\documentclass[sigconf,table,xcdraw]{acmart}

\usepackage{graphicx}
\usepackage{amsmath,amsfonts}
\usepackage{algorithmic}
\usepackage{textcomp}
\usepackage{booktabs}
\usepackage{subcaption}
\usepackage[normalem]{ulem}
\usepackage{indentfirst}
\usepackage{enumitem}

\usepackage{enumitem}
\usepackage{tcolorbox}
\AtBeginDocument{%
  \providecommand\BibTeX{{%
    \normalfont B\kern-0.5em{\scshape i\kern-0.25em b}\kern-0.8em\TeX}}}

\begin{document}

\title{On The Gap Between Software Maintenance Theory and Practitioners' Approaches}

\author{Mivian Ferreira}
\affiliation{%
	\institution{DCC - UFMG}
	\city{Belo Horizonte}
	\country{Brazil}}
\email{mivian.ferreira@dcc.ufmg.br}

\author{Mariza Bigonha}
\affiliation{%
	\institution{DCC - UFMG}
	\city{Belo Horizonte}
	\country{Brazil}}
\email{mariza@dcc.ufmg.br}

\author{Kecia A. M. Ferreira}
\affiliation{%
	\institution{DECOM - CEFETMG}
	\city{Belo Horizonte}
	\country{Brazil}}
\email{kecia@cefetmg.br}


\newcommand{\RQA}{Are developers familiar with the concepts of software metrics, bad smells, refactoring, and change impact analysis?}

\newcommand{\RQB}{Do practitioners apply software metrics, refactoring, bad smells, and change impact analysis in practice?}

\newcommand{\RQC}{Is the background of practitioners associated with the familiarity and the application of software maintenance concepts and techniques?}

\newcommand{\RQD}{Which are the tools most used by practitioners in software maintenance?}

\newcommand{\RQE}{How do practitioners perform change impact analysis?}

\newcommand{\RQF}{Which metrics, refactoring techniques, and bad smells practitioners apply in their activities?}
    
\newcommand{\RQG}{What are the biggest challenges faced by practitioners when carrying out software maintenance?}

\begin{abstract}
The way practitioners perform maintenance tasks in practice is little known by researchers. In turn, practitioners are not always up to date with the proposals provided by the research community. This work investigates the gap between software maintenance techniques proposed by the research community and the software maintenance practice. We carried out a survey with 112 practitioners from 92 companies and 12 countries. We concentrate on analyzing if and how practitioners understand and apply the following subjects: bad smells, refactoring, software metrics, and change impact analysis. This study shows that there is a large gap between research approaches and industry practice in those subjects, especially in change impact analysis and software metrics.
\end{abstract}

\ccsdesc[500]{Software and its engineering~Maintaining software}

\keywords{software maintenance, survey, software engineering, software metrics, refactoring, bad smells, change impact analysis}

\maketitle

\section{Introduction}
\label{sec:introduction}

There is a large gap between industry practice and academic proposals, and this gap tends to grow over time \cite{Parnas:2011}. Some initiatives have been taken for decades to bridge this gap, aiming to comprehend how relevant software engineering research is to practitioners \cite{Lo:2015}\cite{Carver:2016}\cite{Bordin:2018}. There are difficulties for both sides. Research publications might not be accessible to the industry, and their results might not be of practical use \cite{Bern2018}. On the other hand, software engineering researchers may face challenges when collaborating with practitioners, such as differences between culture and time perspective \cite{Runeson2012}.  

We focused this work on the gap between research and practice in software maintenance, one of the most critical and most expensive software live cycle activities. The spectrum of software maintenance subjects is vast and comprises activities such as: log of modification requests, change impact analysis, modification of code and other artifacts, re-engineering, reverse engineering, program comprehension, measurement, migration, tests, training, and daily support \cite{ieee2004_guide}. This work investigates if and how practitioners apply: change impact analysis, software metrics, bad smell, refactoring. We consider those subjects because our research is mostly concentrated on them.

We surveyed 112 practitioners from 12 countries and 92 companies. In particular, we investigate if practitioners are familiar with the subjects considered in this work, if they usually apply such concepts and techniques, the tools regarding those topics practitioners use in practice, and the main challenges developers face when carrying out software maintenance.

\section{Related Work}
\label{sec:realtedwork}

The problem of the distance between what the software engineering research community produces and what practitioners apply in their practices has gained increasing attention \cite{Lo:2015}\cite{Carver:2016}. Murphy-Hill et al. \cite{MurphyHill2012} investigated the adoption of refactoring tools by developers. They found that 90\% of refactoring is performed manually. Their study relies on data of refactoring Java programs in the Eclipse environment. Although popular, Eclipse is only one out of dozens of development environments used nowadays.\footnote{According to the PYPL index, available at https://pypl.github.io/PYPL.html.} We analyze refactoring practice too; however, we did not base our study in any specific language or development environment. Moreover, we considered other three subjects besides refactoring, and we did not gather our data automatically since we performed our analysis by questioning practitioners about their practices.

Our results contradict the assumption that practitioners have widely applied software metrics. Kupiainen et al.'s \cite{Kupiainen2015} systematic literature review on industrial studies analyzed the adoption of software metrics in Agile and Lean software development. They found 102 metrics in the primary studies. However, only one metric is for source code, namely violations of static code, defined as the number of violations found in the static code regarding rules from tools like Findbugs, PMD, and Checkstyle. Dozens of source code software metrics have been proposed, some of them widely known by researchers, such as the CK Metrics. However, none of such metrics appeared in the Kupiainen et al.'s study neither were mentioned by the participants of our research.

Palomba et al. \cite{Palomba2014} investigated the difference between developers' perception and academic concepts of bad smells. Amjed et al. \cite{Amjed2018} investigated bad smells' popularity in Stack Overflow. Both works reached similar conclusions, namely: the academic community might be perceiving bad smell as a problem that is not considered a real problem by developers. The aim and the approach we used are different from those studies since our work investigated if developers know and apply the concept of bad smell. Our results do not contradict the previous ones and empirically show a fundamental reason for the differences in academia and industry perceptions: the bad smell is not a widely known concept among developers.

According to Brodin and Benitti \cite{Bordin:2018}, over 70\% of software developers work with maintenance. However, they point out that it is still a topic little researched. They studied whether practitioners in the industry use the topics about software maintenance taught in undergraduate courses through a survey. They identified that engineering, reverse engineering, software processes, and software measurement are much considered by academia and rarely used in the industry.  In the results reported, Brodin and Benitti state that refactoring is the only topic widely studied by academia that practitioners extensively use. Our results are more precise regarding refactoring since we found that practitioners indeed apply a few refactoring techniques. Moreover, our results indicate that change impact analysis is applied by using mainly naive techniques, such as `search and replace'.

1
All suggestions

\begin{table*}[h]
	\centering
	\caption{Questions and response options of the questionnaire.}
	\label{tab:questionario}
	\resizebox{\textwidth}{!}{%
		\begin{tabular}{p{1cm}p{1cm}p{7cm}p{9cm}}
			\toprule
			\multicolumn{1}{c}{\textbf{Subject}}                           & \multicolumn{2}{c}{\textbf{Question}}                                                                                                                                                                                                             & \multicolumn{1}{c}{\textbf{Reply Options}}                                                                                                                                                                                                                                                                           \\ \midrule
			\multicolumn{1}{c}{\textbf{Challenges to Perform Maintenance}} & \cellcolor[HTML]{C0C0C0}1  & \cellcolor[HTML]{C0C0C0}Describe the main difficulties you face when performing maintenance on software.                                                                                                            & \cellcolor[HTML]{C0C0C0}Open Field                                                                                                                                                                                                                                                                                   \\ \midrule
			& 2                          & Are you familiar with software metrics concept?                                                                                                                                                                      & Yes or No                                                                                                                                                                                                                                                                                                            \\
			& \cellcolor[HTML]{C0C0C0}3  & \cellcolor[HTML]{C0C0C0}What is your opinion about the use of software metrics to ensure the quality of the source code?                                                                                             & \cellcolor[HTML]{C0C0C0}`Very important', `Important', `Little important', `Unnecessary' or `I don't have background to give an opinion.'                                                                                                                                                                            \\
			\multicolumn{1}{c}{\textbf{Software Metrics}}                  & 4                          & Do you use software metrics to evaluate the quality of the source code at your work?                                                                                                                                 & Yes or No                                                                                                                                                                                                                                                                                                            \\
			& \cellcolor[HTML]{C0C0C0}5  & \cellcolor[HTML]{C0C0C0}If you use software metrics to evaluate the quality of the source code at your work, please name them.                                                                                       & \cellcolor[HTML]{C0C0C0}Open Field                                                                                                                                                                                                                                                                                   \\
			& 6                          & If you use metrics to evaluate the quality of the source code at your work, which measurement tool(s) do you use?                                                                                                    & Open Field                                                                                                                                                                                                                                                                                                           \\ \midrule
			& \cellcolor[HTML]{C0C0C0}7  & \cellcolor[HTML]{C0C0C0}Are you familiar with the concept of refactoring ?                                                                                                                                           & \cellcolor[HTML]{C0C0C0}Yes or No                                                                                                                                                                                                                                                                                    \\
			& 8                          & Have you ever applied code refactoring at your work?                                                                                                                                                                 & Yes or No                                                                                                                                                                                                                                                                                                            \\
			\multicolumn{1}{c}{\textbf{Refactoring}}                       & \cellcolor[HTML]{C0C0C0}9  & \cellcolor[HTML]{C0C0C0}If you have ever used code refactoring at your work, what kind (s) of refactoring did you use?                                                                                               & \cellcolor[HTML]{C0C0C0}Open Field                                                                                                                                                                                                                                                                                   \\
			& 10                         & If you have ever used code refactoring at your work, have you used a tool for this?                                                                                                                                  & Yes or No                                                                                                                                                                                                                                                                                                            \\
			& \cellcolor[HTML]{C0C0C0}11 & \cellcolor[HTML]{C0C0C0}If you have ever used code refactoring at your work and have used a tool to do so, which tool (s) did you use?                                                                               & \cellcolor[HTML]{C0C0C0}Open Field                                                                                                                                                                                                                                                                                   \\ \midrule
			& 12                         & Are you familiar with the concept of bad smell?                                                                                                                                                                      & Yes or No                                                                                                                                                                                                                                                                                                            \\
			\multicolumn{1}{c}{\textbf{Bad Smell}}                         & \cellcolor[HTML]{C0C0C0}13 & \cellcolor[HTML]{C0C0C0}When developing or maintaining a system at work, do you usually check bad smells in the source code?                                                                                         & \cellcolor[HTML]{C0C0C0}Yes or No                                                                                                                                                                                                                                                                                    \\
			& 14                         & If you answered 'yes' to the previous question, what are the bad smells most commonly detected by you?                                                                                                               & Open Field                                                                                                                                                                                                                                                                                                           \\ \midrule
			& \cellcolor[HTML]{C0C0C0}15 & \cellcolor[HTML]{C0C0C0}Have you ever noticed whether a change performed in a software system by you had caused the need to make other changes not initially foreseen?                                               & \cellcolor[HTML]{C0C0C0}`Never', `Few times', `Oftentimes' or `Always'                                                                                                                                                                                                                                               \\
			& 16                         & Are you familiar with the term ”Change Impact Analysis ”?                                                                                                                              & Yes or No                                                                                                                                                                                                                                                                                                            \\
			\multicolumn{1}{c}{\textbf{Change Impact}}                     & \cellcolor[HTML]{C0C0C0}17 & \cellcolor[HTML]{C0C0C0}When correcting a bug (error or failure), performing a change or creating a new functionality in the system, do you usually analyze the impact of the change in the rest of software system? & \cellcolor[HTML]{C0C0C0}Yes or No                                                                                                                                                                                                                                                                                    \\
			& 18                         & What kind of technique do you apply to analyze parts of the software that need to be modified?                                                                                                                       & `I explore the code manually and intuitively, not always with prior knowledge about it.', `I explore the code manually guided by the prior knowledge I have about it.', `I use a tool for this.', or `I do not analyze all the parts that need to be modified, I make the modifications as I identify the problems.' \\
			& \cellcolor[HTML]{C0C0C0}19 & \cellcolor[HTML]{C0C0C0}If you use a tool to analyze which parts of the software need to be modified, please name them.                                                                                              & \cellcolor[HTML]{C0C0C0}Open Field                                                                                                                                                                                                                                                                                   \\ \bottomrule \\
		\end{tabular}%
	}
\end{table*}

\section{Study Design}
\label{sec:studydesign}

This section describes the study's design presented in this paper, detailing the following aspects: the research questions, the construction and validation of the questionnaire, the participants' selection, and the method we applied to analyze the data.

\subsection{Research Questions}
\label{sec:rq}

The research questions aim to elucidate if and how some of the main concepts and techniques proposed by the research community for software maintenance are applied. This section presents the research questions and how we analyzed the data to answer them.

\vspace{0.1cm}
\noindent  \emph{RQ1. \RQA} 

To answer this question, we calculated the percentage of the participants who answered `yes' to the questions related to familiarity (Rows 2, 7, 12, and 16 of Table \ref{tab:questionario}). We used a yes/no question to split the respondents into two distinct groups \cite{StaCan}, once we want to know if (not how) the respondent is familiar with the concepts.\par

\vspace{0.1cm}
\noindent \emph{RQ2.\RQB} 

To answer this research question, we calculated the percentage of the participants who answered `yes' to the questions described in Rows 4, 8, 13, and 17 of Table \ref{tab:questionario}. For the same reasons mentioned in RQ1, we also chose to use a yes/no question in this case.\par

\vspace{0.1cm}
\noindent \emph{RQ3. \RQD} 

In this research question, we aim to identify which tools are used to support the studied topic's activities. The participants answered the questions associated with RQ3 in a text field. To analyze the data, we read all the answers and tabulated the tools described by the participants.\par

\vspace{0.1cm}
\noindent \emph{RQ4. \RQE} 

With this research question, we investigate if and how practitioners analyze the impact of changes they need to perform in software systems. To answer this question, we summarized and reported the answers to the question described in Row 18 of Table \ref{tab:questionario}.\par

\vspace{0.1cm}
\noindent \emph{RQ5. \RQF} 

To answer this research question, we read all the answers given to the questions described in Rows 5, 9, and 14 of Table \ref{tab:questionario} and summarized the data.\par

\vspace{0.1cm}
\noindent \emph{RQ6. \RQG} 

To investigate this research question, two authors of this work tabulated and labelled, separately, the answers to the question shown in the first row of Table \ref{tab:questionario}. In the presence of different labels for the same answers, these two authors opted for the final label by a consensus between them.

\subsection{Questionnaire Construction}
\label{questionnaire}

We based the questionnaire on the guideline of Kitchenham and Pfleeger \cite{Kitchenham:2008}. It is composed of seven sections described as follows. Table \ref{tab:questionario} shows the survey questions and answers options. 

\noindent \emph{Term of consent} introduced the study's purpose to the participants and requested their endorsement to use the data collected and participants' anonymity assurance.

\noindent \emph{Participants' characterization.} Collect data about the participants' professional lives: (i) name and country of the company where they work; (ii) the position held in the company; (iii) academic background; (iv) years of professional experience; (v) programming languages currently used in their jobs; and (vi) the methodology used in their company's software development process.

\noindent \emph{Challenges to perform software maintenance.} This section contains only one question in which we requested the participants to describe the main challenges and difficulties they face to perform maintenance activities. 

\noindent \emph{Metrics.} In this section, we asked the participants if they are familiar with metrics; and consider them useful. We also asked if they use metrics to measure the code quality, the tools they use to do it, and the most common metrics they apply.

\noindent \emph{Refactoring.} About refactoring, we asked if the participants are familiar with the term ``refactoring'' and if they usually perform code refactoring. We also asked them to name the types of refactoring techniques they apply and their tools to perform such activity.

\noindent \emph{Bad smells.} We asked if the participants are familiar with the term ``bad smell'', and if they use it to verify the presence of bad smells in the software code. If so, we requested them to list which bad smells they use to search in software systems. 

\noindent\emph{Change impact analysis.} We asked the participants: if they notice the need of changing other pieces of code when they carry out a specific change in the code; if they are familiar with the term ``change impact analysis''; if they use to search other pieces of code that need to be modified when they make a change in the code; what kinds of techniques they use to perform such analysis; and, finally, if they use any tool to perform change impact analysis.

\subsubsection{Validation of the Questionnaire}

Before sending the questionnaire to the participants, we made a pilot survey to test the questionnaire and identify improvement needs. For this purpose, we sent the questionnaire to two developers from different companies. We identified the following primary need for improvements from the first version. (1) Observing the answers, we notice that the questions we asked before the section ``Challenges to Perform Software Maintenance''may had influenced the participants' answers about the main challenges they face in software maintenance. Probably, this happens because we asked about refactoring, dependency analysis, bad smells, and metrics before that section, which may have induced the participants to include problems related to such subjects in their answers. We then decided to put the Section `Challenges to Perform Software Maintenance' as the first one in the questionnaire. (2) We reformulated the answer options that involve ranges: number of employees and years of professional experience. (3) We divided some long questions into two or more to be more precise and improve the data analysis. 

We sent the second version of the questionnaire with such improvements to eight developers in a second round. We made a previous analysis of the responses to ensure that the questionnaire was more precise and able to gather the information we need. Then, we sent the survey to the other participants.

\subsection{Participants Selection}

We invited practitioners to answer the survey in two ways: directly and indirectly. Some participants were contacted directly via email, LinkedIn, and Facebook and were chosen by convenience since the authors had their contact. To motivate the practitioners to answer the questionnaire, we chose to send a particular message to each participant explaining the survey's aim and inviting him/her to participate in the research. In this message, we asked them to forward the email to other colleagues, aiming to invite more participants. We also published the questionnaire in our social networks and specialized IT mailing lists.

We received 112 responses. We directly contacted 204 practitioners; out of this amount, 77 answered the questionnaire, achieving a response rate of 37.8\%. We reached the remaining 35 practitioners that answered the questionnaire indirectly. As we aimed to have a global assessment of practitioners' perceptions and practices worldwide, we tried to contact professionals from different countries and many companies. The practitioners that answered the questionnaire are from 92 companies and 12 countries.


\section{Participants Characterization}
\label{sec:participantscharacterization}

This section describes the characteristics of the participants. 
\begin{figure}[!h] 
    \centering
    \includegraphics[trim=0 15 0 55,clip,width=0.9\linewidth]{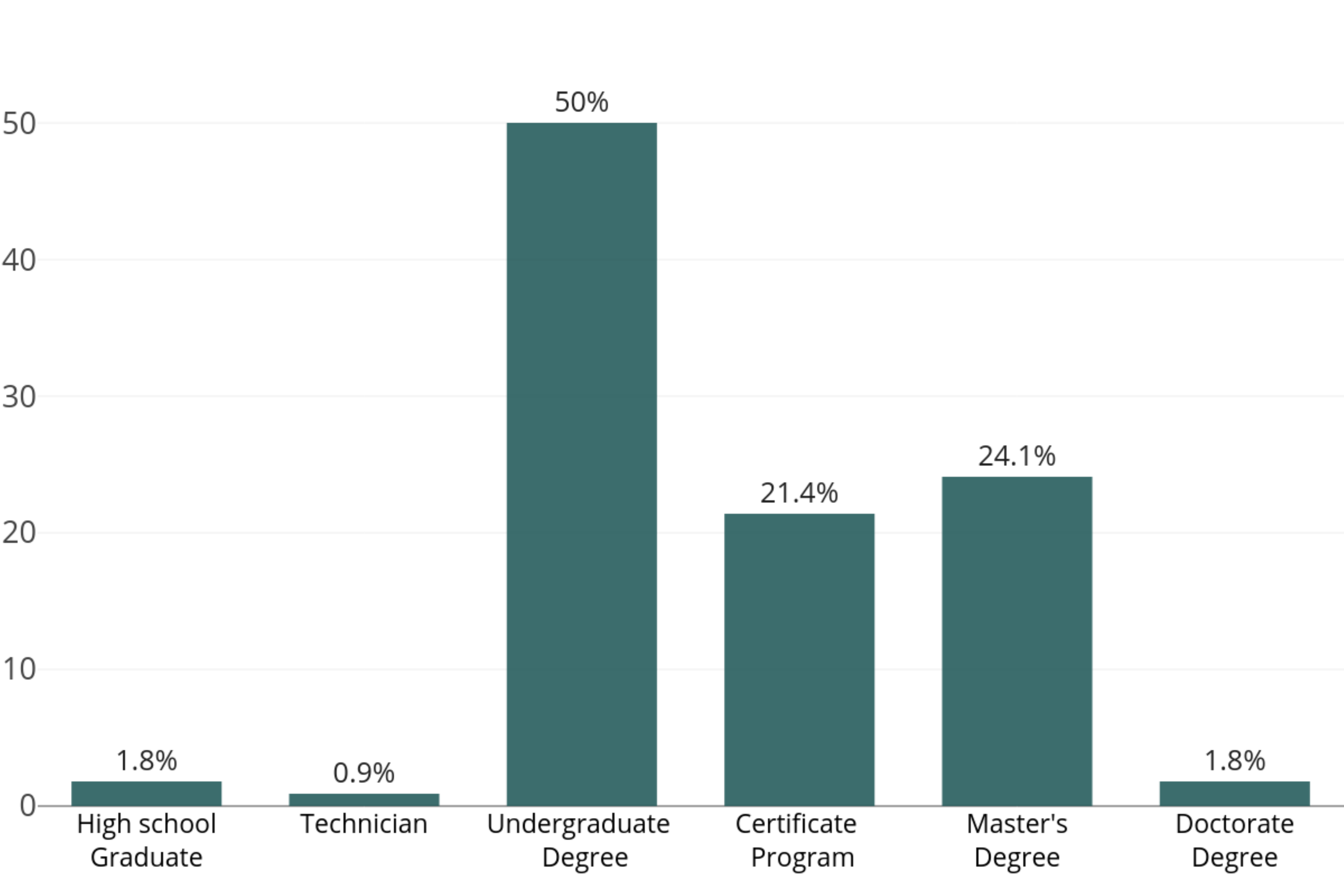}
    \caption{Distribution of participants' academic background.}
    \label{fig:escolaridade} 
\end{figure}

\noindent \textit{Academic Background.} The participants with an undergraduate degree, certificate program, or a master's degree correspond to 95.5\% of the sample, as shown in Figure \ref{fig:escolaridade}. Only 4.5\% of the participants are high school graduates, technicians, or have a Ph.D. degree, which indicates that the practitioners have a high level of formal education, i.e., they had access to Computer Science's theoretical aspects.

\begin{figure}[!h] 
    \centering
    \includegraphics[trim=0 10 0 30,clip,width=0.9\linewidth]{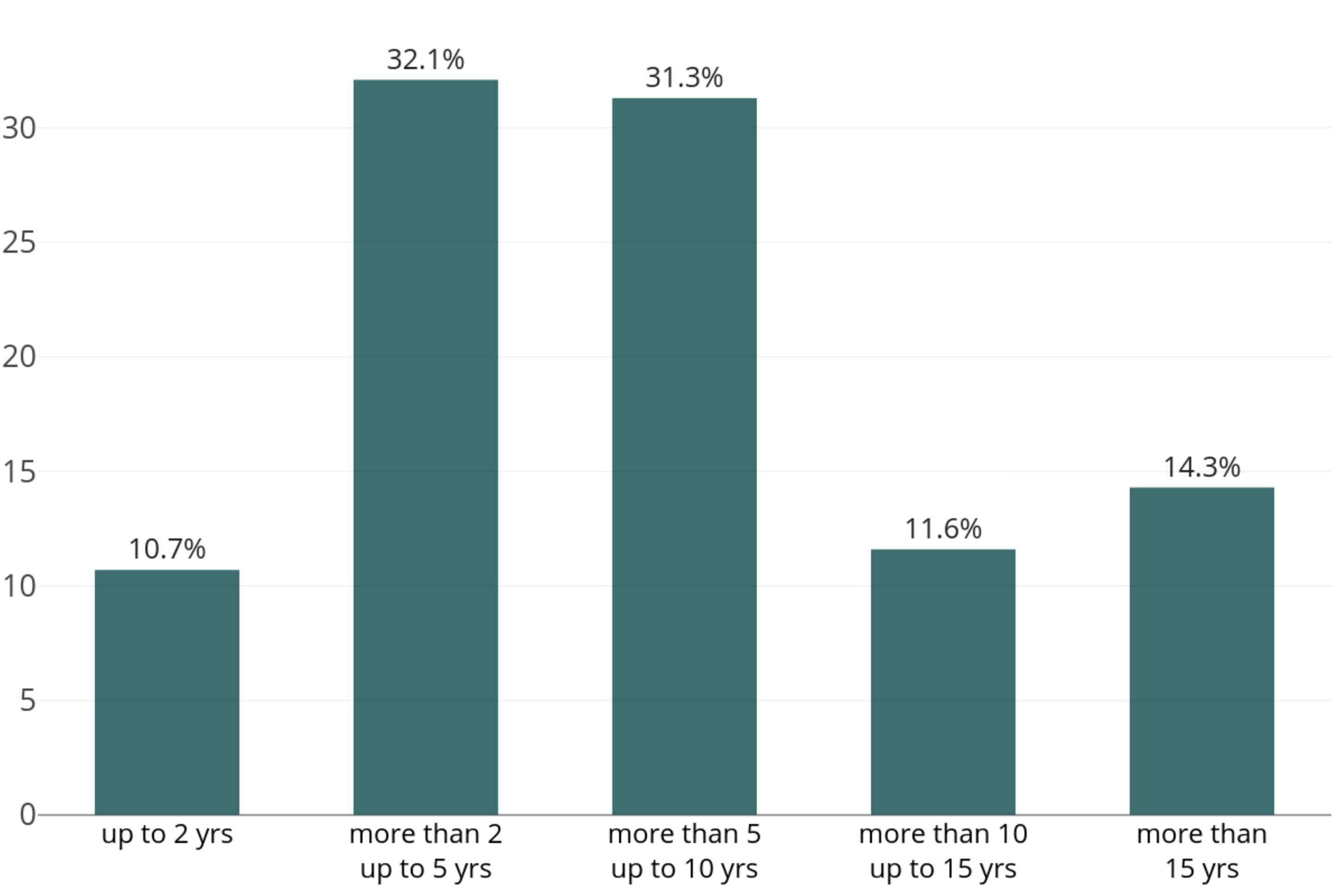}
    \caption{Distribution of participants' professional experience.}
    \label{fig:experiencia} 
\end{figure}

\noindent\textit{Professional Experience.} Most participants, 63.4\%, have between two and ten years of professional experience, and 25.9\% of the participants are experienced professionals; they have more than ten years of career. Junior practitioners are 10.7\% of the participants. Figure \ref{fig:experiencia} shows the distribution of these data. Therefore, most participants (89\%) have much practical experience and are likely to know well real software engineering scenarios.

\noindent\textit{Programming Languages.} The programming languages used by the participants are Java, C\#, C++, Python, JavaScript, PHP, Scala, Kotlin, TypeScript, ShellScript, Delphi, Swift, Objective C, Golang (Go), Groovy, Pearl, Dart, Ruby, Visual FoxPro, VB.NET, and ASP.NET.

\noindent\textit{Methodologies.} Agile methodologies are the most widely used by developers - being Scrum and XP the most cited of them. Only one participant informed the company uses a Waterfall process.

\noindent\textit{Companies' Sectors.} 71\% of the companies are from the IT area. The other companies are in the following areas: trade, financial, bank, industry, marketing, health, education, and government.

\begin{figure}[!h] 
    \centering
    \includegraphics[trim=0 20 0 55,clip,width=0.95\linewidth]{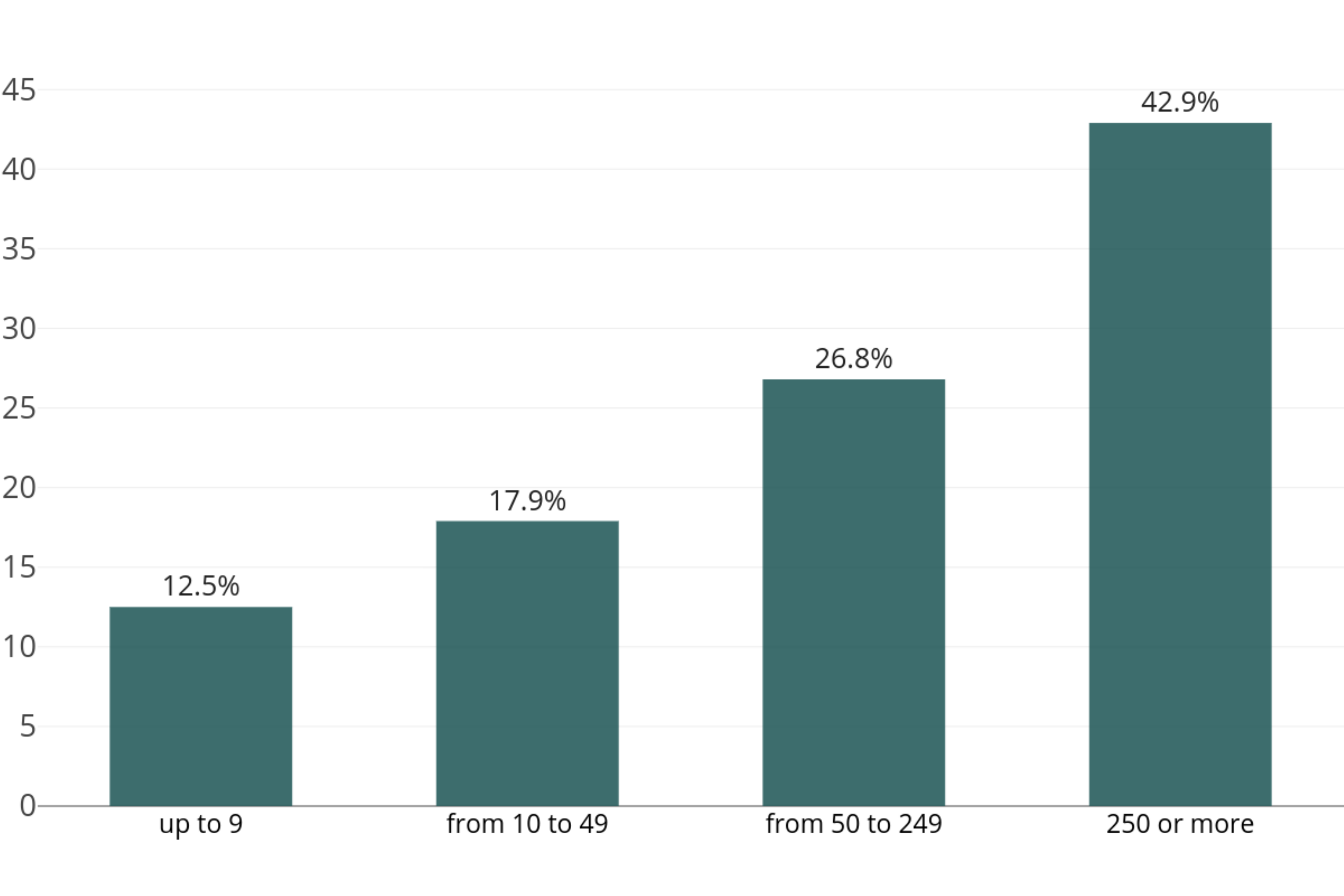}
    \caption{Distribution of participants' companies size by the number of employees.}
    \label{fig:empresa} 
\end{figure}

\noindent\textit{Companies' Characterization.} Among the participants, 69.7\% work in medium-sized or large companies, and 30.3\% work in micro and small companies. Figure \ref{fig:empresa} presents the distribution of the participants' companies size by the number of employees.

\section{Results}
\label{sec:results}

This section presents the data analysis and the answers to the research questions.

\subsection*{\large RQ1. \RQA}

\begin{figure}[h] 
    \centering
    \includegraphics[trim=0 35 0 90,clip,width=0.85\linewidth]{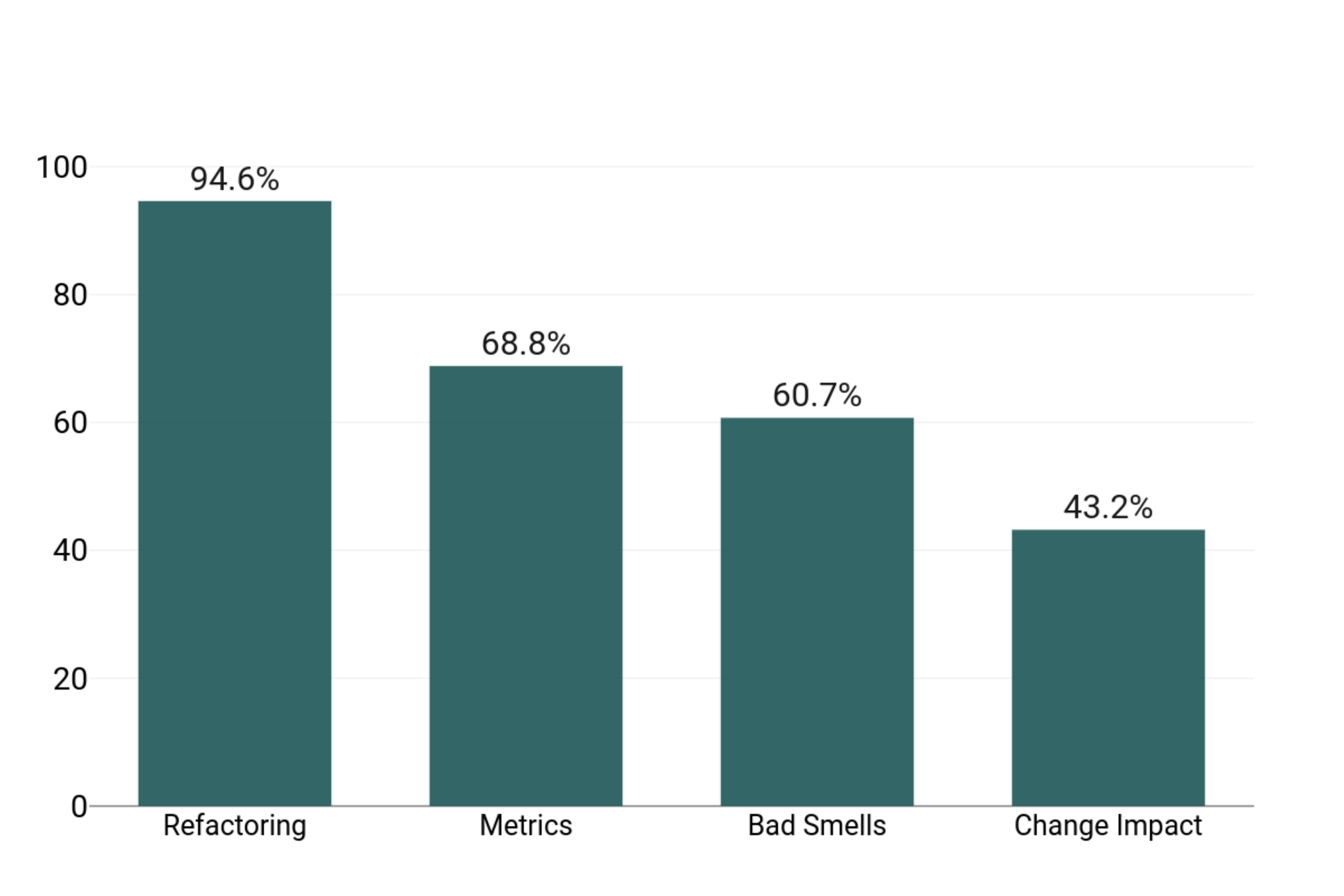}
    \caption{Percentage of participants who declare to be familiar to the subjects.}
    \label{fig:familiar} 
\end{figure}

To answer this question, we considered the number of participants who declared to be familiar with the subjects investigated in this paper. Figure \ref{fig:familiar} shows the percentage of participants who answered `yes' to the questions regarding the familiarity with software metrics, refactoring, bad smells, and change impact analysis, respectively (see Rows 2, 7, 12, and 16 of Table \ref{tab:questionario}).

Refactoring is the most popular subject among the participants; 94.6\% of them claim to be familiar with refactoring. Metrics is the second most popular subject; 68.8\% of the participants are familiar with it. Bad Smell is a term known by 60.7\% of the participants. Change Impact Analysis is the least familiar term to the practitioners who answered the survey, 43.2\%, stated they are familiar with this concept.

\subsection*{\large RQ2. \RQB} 

\begin{figure}[h] 
    \centering
    \includegraphics[trim=0 20 0 55,clip,width=0.8\linewidth]{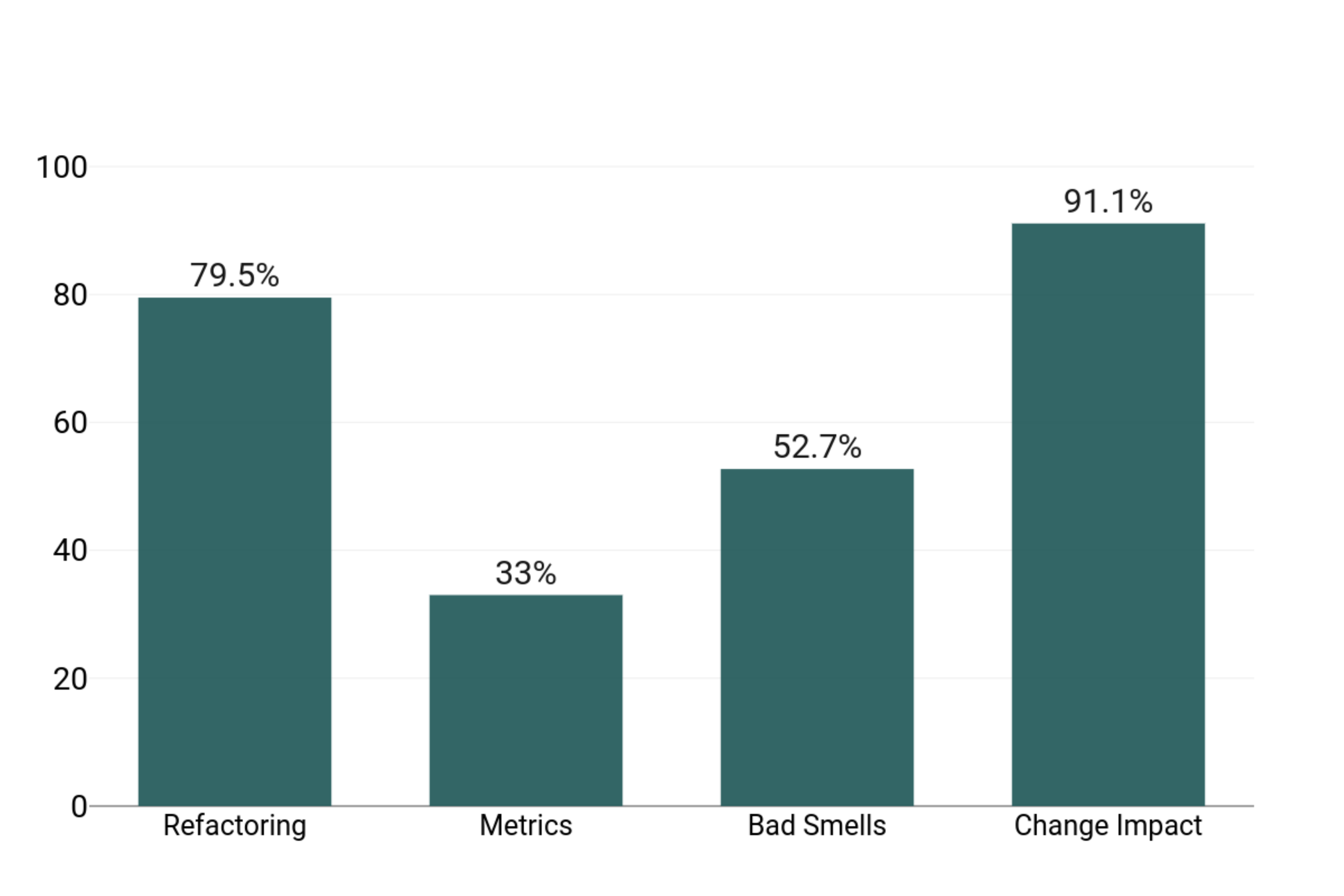}
    \caption{Percentage of participants who declared to apply refactoring, software metrics, bad smell, and change impact analysis.}
    \label{fig:aplicacao} 
\end{figure}

To answer RQ2, we calculated the number of participants that answered `yes' to the questions regarding the practical application of metrics, refactoring, bad smells, and change impact analysis (Rows 4, 8, 13, and 17 of Table \ref{tab:questionario}). Figure \ref{fig:aplicacao} shows the results.

\noindent\textit{Change impact analysis} is the technique most applied by practitioners, 91.1\% of the participants affirmed that they perform some extent of change impact analysis when making software code modifications. It is important to note that only 43,2\% informed they are familiar with the concept of ``change impact analysis''. We intentionally used the term ``change impact analysis'', which is commonly used in academia, when asking if the participant is familiar or not with the concept. However, we did not use this term in the other questions about performing change impact analysis. We constructed the questionnaire in such a way because, in this specific case, we assumed that even not knowing the terminology used by researchers, the participants may use any change impact analysis in practice. This result shows that the industry did not well adopt the researchers' jargon on this topic. 

\noindent\textit{Refactoring} is the concept that the participants are most familiar with (94\%). Besides, 79.5\% of the participants perform code refactoring. \textit{Bad smell} is a more balanced concept in terms of theoretical knowledge and practice: 60.7\% declared to know the concept of bad smell, and 52.7\% of the participants affirmed to verify bad smells in software code. 

\noindent\textit{Software metrics} are the second most popular concept (68,8\%). Nevertheless, it is the least applied one. Only 33\% of the participants declared to use software metrics. However, 68.8\% of the participants consider software metrics essential or very important, contrasting with 9.8\% who believe that it is of little importance; 21.4\% informed they do not have the background to manifest their opinion about such subject. These results suggest that when knowing the concept of software metrics, the practitioner is likely to consider their application important but not always actually apply it.

\subsection*{\large RQ3. \RQD} 

We asked participants whether they use tools to collect metrics, perform refactoring, and change impact analysis. In the sequence, we present the results for each of the subjects covered.

\begin{figure}[!h] 
    \centering
    \includegraphics[trim=10 30 40 60,clip,width=0.8\linewidth]{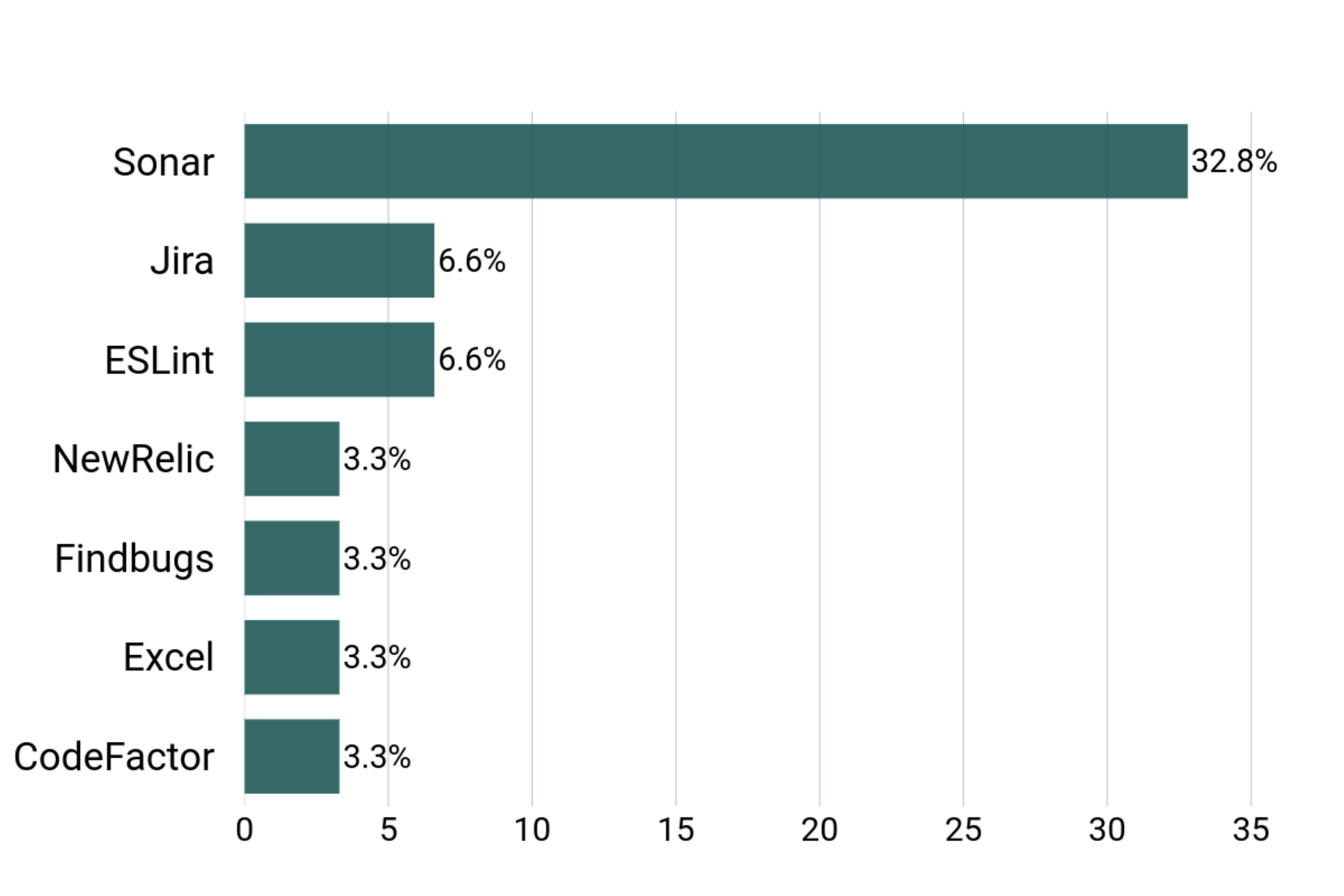}
    \caption{Tools most used by practitioners to collect metrics.}
    \label{fig:ferramentas_metricas} 
\end{figure}

\noindent \textit{Software Metrics.} Only 33\% of the participants affirmed to use a software measurement tool. The participants pointed out 62 tools. Figure \ref{fig:ferramentas_metricas} shows the citation percentage of the most used tools. SonarQube is the most used tool; 32.8\% of practitioners used this platform to collect metrics. About 6.6\% of the participants cited ESLint and Jira, and 3.3\% of practitioners pointed out CodeFactor, Excel, FindBugs, and NewRelic. Practitioners mentioned other tools just once.

\begin{figure}[h] 
    \centering
    \includegraphics[trim=10 30 40 60,clip,width=0.8\linewidth]{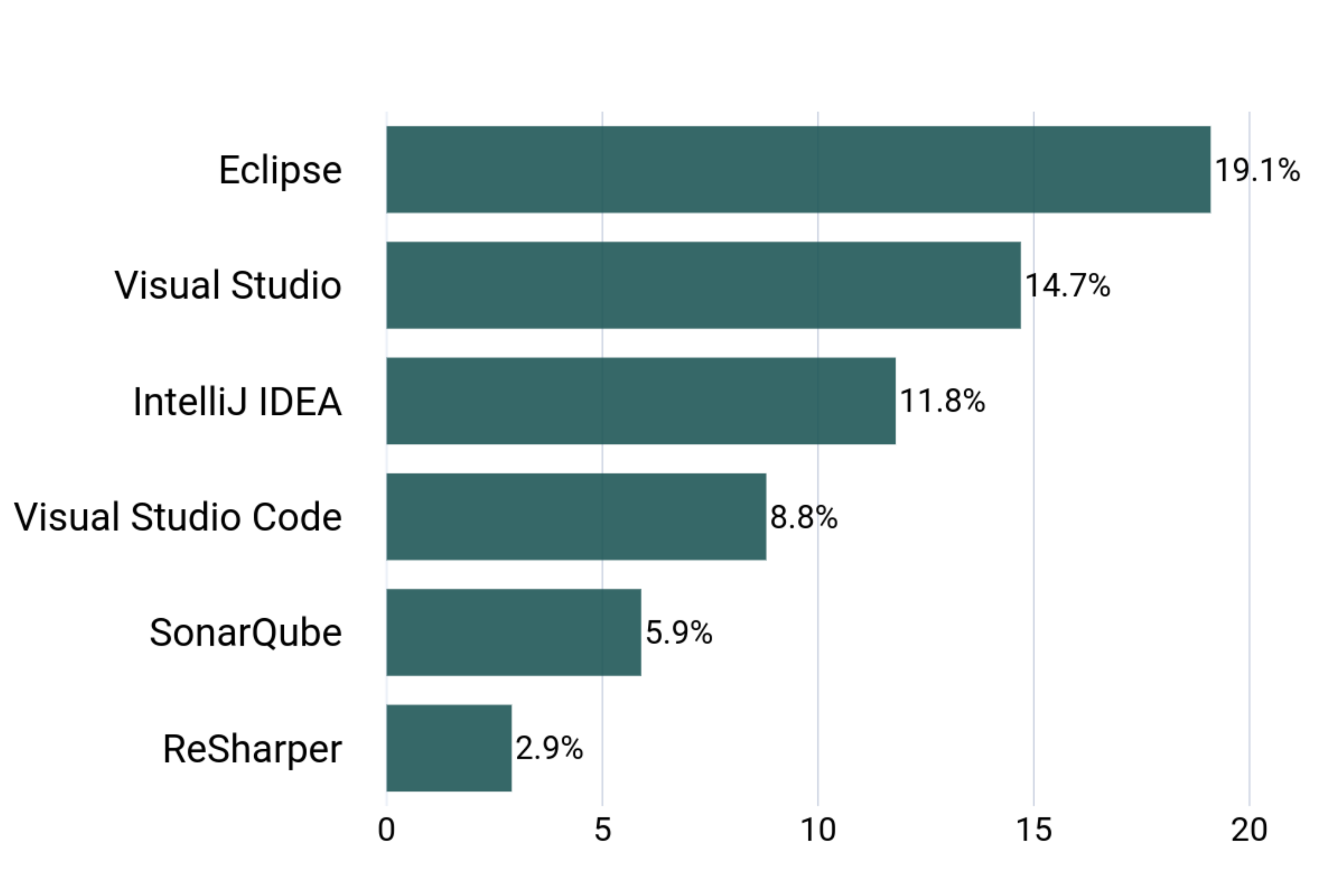}
    \caption{Tools most used by practitioners to perform refactoring.}
    \label{fig:ferramentas_refactoring} 
\end{figure}

\noindent \textit{Refactoring.} Only 36.6\% of the participants declared they use tools to perform refactoring. They mentioned 68 tools. It is worthwhile to notice that 79.5\% of the participants perform refactoring. Therefore, we may assume that they perform refactoring manually. Figure \ref{fig:ferramentas_refactoring} shows the distribution of the citations of the tools. The most commonly used tools for refactoring are the IDE Visual Studio (37.2\%), Eclipse (30.2\%), and IntelliJ IDEA (11.8\%), or an extension of an IDE, such as ReSharper (4.7\%), an extension of the Visual Studio platform. 9.3\% of the participants use SonarQube to perform code refactoring. Participants mentioned the other tools just once.

\begin{figure}[h] 
    \centering
    \includegraphics[trim=10 30 40 60,clip,width=0.8\linewidth]{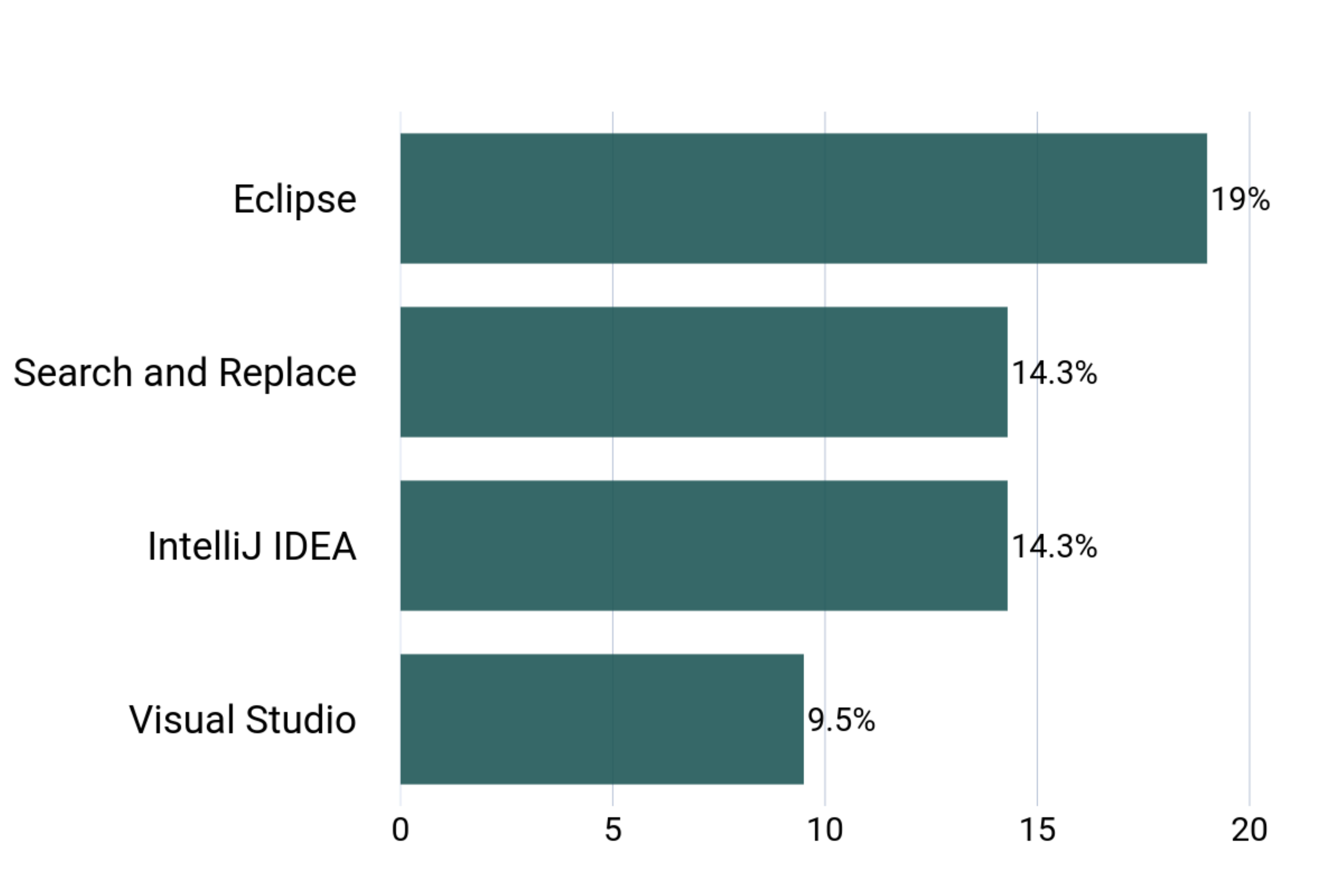}
    \caption{Tools most used by practitioners to perform change impact analysis.}
    \label{fig:ferramentas_change_impact} 
\end{figure}

\noindent \textit{Change Impact Analysis.} The participants pointed out 13 tools they use to perform this task. Figure ~\ref{fig:ferramentas_change_impact} shows the percentage of the most used tool of change impact analysis. In this case, IDE is also the most cited tool: Eclipse, IntelliJ IDEA, and Visual Studio. It is essential to mention that 14.3\% of the participants mentioned preferring text search (``search and replace'') to perform change impact analysis, and this was the second most cited approach. However, none of them is a specific tool for change impact analysis.

\subsection*{\large RQ4. \RQE}

\begin{figure}[!ht] 
    \centering
    \includegraphics[trim=0 20 0 60,clip,width=\linewidth]{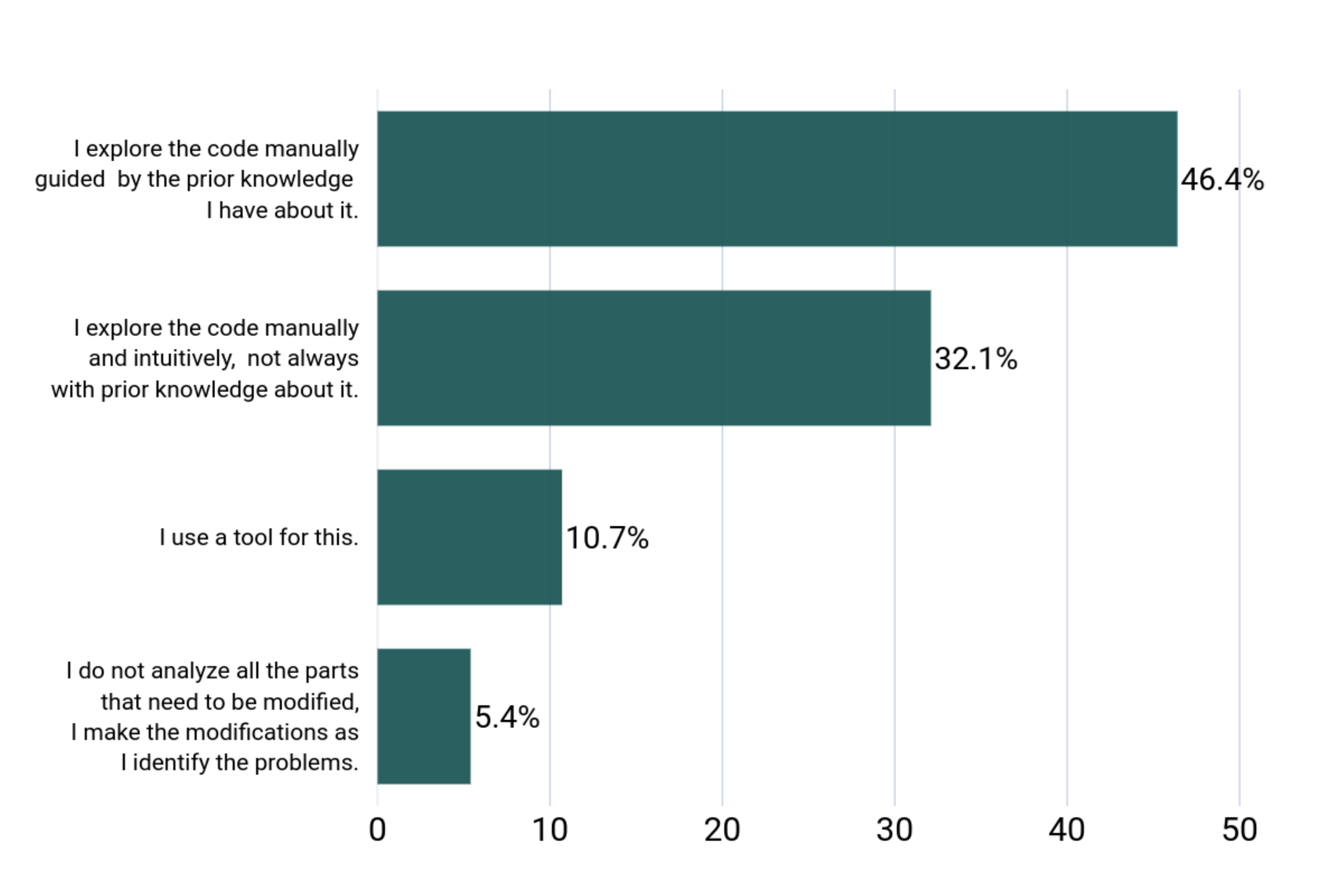}
    \caption{How practitioners perform change and impact analysis.}
    \label{fig:change} 
\end{figure}

In this research question, we investigate whether and how practitioners perform change impact analysis. Figure~\ref{fig:change} shows the rank of the participants' answers.

Most of the participants, 46.4\%, informed that they analyze parts of the code that need to be modified by exploring the code manually and intuitively, guided by prior knowledge about the software source code. Another significant part of the participants, 32.1\%, declared that they explore the code manually and intuitively, not always based on prior knowledge about the system. Only 10.7\% of the participants pointed out the use of tools to assist in change impact analysis. A small part of the participants, 5.4\%, declared that they do not analyze all the elements that need to be modified and make the modifications as they identify the problems. Other ways of performing change impact analysis correspond to 3.6\% of the responses.

\subsection*{\large RQ5. \RQF}

We asked the participants to point out the most common software metrics they use, the commonly refactoring techniques performed by them, and the bad smells they use to consider.

\noindent \textit{Refactoring.} The most common refactoring techniques applied by the participants are: \textit{Extract Method} (21.43\%), \textit{Rename Method} (13.39\%), and \textit{Extract Class} (12.5\%). 

\noindent\textit{Metrics.} The analysis of RQ1 - \textit{\RQA} - showed that software metrics is the second well-known subject. However, the participants just pointed out a few software metrics. The most cited terms regarding software metrics were: \textit{Number of Bugs} (9.9\%); \textit{Test Coverage} (8.91\%); and \textit{Cyclomatic Complexity} (7.92\%). In the last decades the literature proposed many software metrics. Nevertheless, the results of this survey show that they have not been widely applied in the industry. In particular, we noticed that practitioners do not mention the widely known software metrics in academia, such as those proposed by Chidamber and Kemerer \cite{Chidamber:1994}.

\noindent \textit{Bad Smell.} The most cited bad smells were \textit{Duplicate Code} (23.21\%), \textit{Long Method} (19.64\%), and \textit{Long Class} (9.82\%). This result indicates that developers' mainly assess code structure quality by means of code duplication, method size, and class size.

\subsection*{\large RQ6. \RQG}

\begin{figure}[!ht] 
    \centering
    \includegraphics[width=\linewidth]{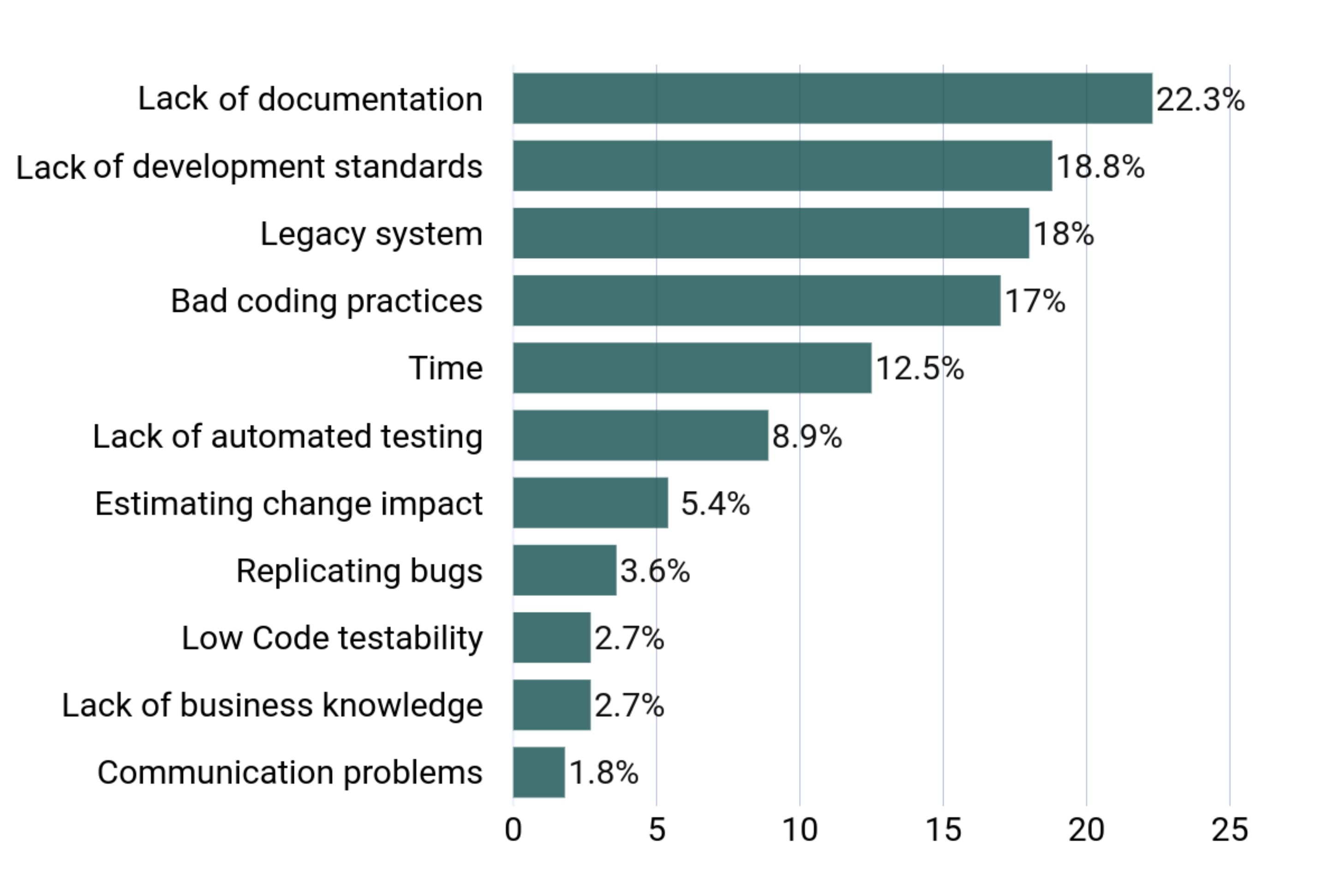}
    \caption{Most important challenges faced by developers in software maintenance.}
    \label{fig:desafio} 
\end{figure}

We asked the participants to write in an open text field, which are the main difficulties they face when performing maintenance on the software systems. The participants indicated several challenges in software maintenance. We reported them in Figure \ref{fig:desafio}.

\noindent \textit{Lack of documentation} is the most cited problem; 22.3\% of the participants have this perception. In general, they described that this problem is made more critical due to the high turnover, which is very common in IT companies.

\noindent \textit{Lack of standard for software development,} is cited by 18.8\% of the participants and is the second biggest challenge. The participants mentioned that the companies' development patterns are not always followed by developers, making the code difficult to understand and change.

\noindent \textit{Legacy system} is cited by 18\% of the participants as a challenge in software maintenance. In this context, the participants refer to the legacy system as ``\textit{long-standing codes in which several people have already worked}''.

\noindent\textit{Bad coding practices} are cited by 17\% of the participants. According to the participants, this problem mainly involves low code readability, poorly structured functions, replicated methods in various parts of the code, and non-explanatory comments.

\noindent\textit{Time} is also mentioned (12.5\%). The participants' main issues are the short deadlines for performing maintenance-related activities and little time to understand the code and implement the change.

\noindent\textit{Lack of automated testing} (8.9\%), difficulty in \textit{Replicating Bugs} (3.6\%), and \textit{Low Code Testability} (2.7\%) are also challenges raised by the participants. The participants described that the absence of unit, regression, and integration tests negatively affects software code regarding automated testing. The participants associate the difficulty in code testability with the maintenance of monolithic and poorly structured systems.

\noindent\textit{Estimating change impact} is pointed out as a challenge by 5.4\% of the participants. According to them, not knowing how a change in the code will impact the software system raises the system's ``fear'' of side effects.

\noindent\textit{Lack of business knowledge} (2.7\%) and \textit{Communication problems} (1.8\%) also appeared as challenges for software maintenance. The participants reported that when the customer does not have a solid knowledge of the business, they demand more system changes. Practitioners also pointed out that the problems of communication with customers and coworkers negatively impact software maintenance.

\section{Discussion}
\label{sec:discussion}

The results of this study lead to some insights, which we discuss in this section.

\subsubsection*{\textbf{\emph Software metrics are not fully applied in practice}.}

\noindent The source code software metrics mostly considered in the literature, are not applied in practice. Some possible causes for that should be investigated: lack of proper tools, lack of thresholds for the metrics, lack of knowledge of the source code metrics by developers, or the metrics proposed in the literature are not of practical use. The most used metrics are the number of bugs, test coverage, and cyclomatic complexity.

\subsubsection*{\textbf{\emph Refactoring is a popular concept, but only the simple refactoring techniques.}}

The refactoring techniques performed in practice are simple and provided by IDE: \textit{Extract Method}, \textit{Rename Class}, and \textit{Extract Class}. It is essential to note that 93.7\% of participants indicated that their companies use agile methodologies or a mix of agile methodologies. Among those methodologies, the participants indicated XP (eXtreme Programming), refactoring as the primary practice. Therefore, the popularity of the agile methodologies might be the cause of the popularity of refactoring. Moreover, agile methodologies emerged from the industry and not academia, which may also explain its popularity. It is essential to investigating the reasons why developers do not commonly apply other refactoring types. 
 
\subsubsection*{\textbf{\emph The bad smell issue.}}

Although bad smell is known by 60.7\% of the respondents, only 52.7\% apply the concept in practice. Previous work have pointed out two main possible reasons for that: what is considered as a potential design problem by researchers might not be a real problem in practice; bad smells detection is not of practical use. We add two hypotheses point out by the participants: developers do not apply bad smells because they do not know the concept deeply; lack of proper tools and strategies for bad smell detection.

\subsubsection*{\textbf{\emph Change impact analysis is not adequately performed in practice.}}

The results concerning change impact analysis showed that even not knowing the term used in the literature, it is the most applied technique by the practitioners. A possible cause is that change impact analysis is an intrinsic and indispensable activity in software maintenance. Nevertheless, practitioners do not use the proper tools to perform it. Most developers (78.5\%) perform change impact analysis manually, guided or not by the previous knowledge they have about the code analyzed, as described in the results. The participants reported they still use inadequate mechanisms such as ``search and replace'' to perform change impact analysis. Additionally, the participants pointed out change impact analysis as a significant challenge in software maintenance. These results indicate the need for proper techniques to aid change impact analysis.

\subsubsection*{\textbf{\emph Difficulties source code maintenance.}}

The main challenges in performing code maintenance are lack of documentation, lack of standard software development, legacy systems, and bad coding practices. The research community well knows such problems. However, this result indicates that the proposed solutions for such problems were not enough to overcome such challenges in practice despite the effort made to solve them.

\subsubsection*{\textbf{\emph Challenges for academia and industry}}

The results indicate that the gap between academia and industry in software maintenance is large. For instance, refactoring and bad smell are related concepts first published in 1999 \cite{Fowler:}. The knowledge of refactoring has spread, but the knowledge of bad smell has not. Moreover, practitioners know only simple refactoring techniques. Object-oriented software metrics have been discussed since the early '90s but were not adopted by the practitioners, although they consider software measurement necessary. Change impact analysis is an intrinsic and challenging task of software maintenance and has been studied by researchers. However, the industry does not apply any efficient automatic tool in this task. We may raise some non-exclusive hypotheses for that: some techniques proposed in academia is not useful in practice; the lack of proper methods, tools, and recommending systems for detecting bad smells, change impact analysis, and software measurement may be a reason why such techniques are not well applied in practice; the software engineer education may be outdated; the way the academic proposals are published does not reach practitioners. The results emphasize that the research community should strive to define more proper techniques and tools to aid software maintenance practice. On the other hand, the industry should invest more in updating the knowledge on the field.

\section{Threats to Validity}
\label{sec:ameacas}

We based the survey data on the responses collected by a questionnaire. Therefore, the data are susceptible to the participants' interpretation. To mitigate this issue, we ran a pilot questionnaire with accurate data for our analysis. Based on the answers collected in this initial round, we constructed a final survey to remove ambiguities and biases.

Our study relies on the responses given by 112 participants. We just considered exclusively participants that are professionals of software development and maintenance. Besides, the sample used in this paper is composed of participants from 92 companies and 12 countries, with a wide range of years of professional experience, with all kinds of academic backgrounds, and using a large number of programming languages. 

To identify the main challenges the participants face when performing software maintenance, we asked them to describe their difficulties in an open text field. The answers to this question were manually categorized. Therefore, they are subject to interpretation by those who performed this categorization.  To mitigate this threat to validity, we standardized the labels used in the categorization, i.e., we identified the keywords mentioned in the participants' responses, such as documentation, legacy system, readability, and standard, among others.  Besides, the label assigned to each answer was made separately by two authors of this paper. After this, we compared the classifications to obtain the final classification of each answer. In case of divergence in the categorization, both authors analyzed it to obtain a consensus.

\section{Conclusion}
\label{sec:conclusion}

 We surveyed software practitioners to investigate whether and how software maintenance techniques have been applied in practice. In particular, we investigated the usage of the following concepts and techniques: software metrics, refactoring, bad smells, and change analysis impact. For this purpose, we surveyed 112 software development practitioners from 92 companies and 12 countries.
The results showed that change impact analysis is the most applied technique among the ones considered in this work. However, there is a lack of proper tool support to perform change impact analysis. Although refactoring is a widespread technique, few refactoring techniques have been applied in practice. Moreover, refactoring is mostly provided by IDE. Bad smells and software metrics are the less known and applied concepts. 

This study also revealed that participants considered the lack of system documentation, lack of development patterns, and legacy software as the leading software maintenance challenges. The results indicate that software maintenance demands even more community effort to develop and provide proper tools and methods for software maintenance, especially in change impact analysis and software measurement. 

We intend to perform two additional analyses with this survey's data: the differences in how practitioners approach software maintenance depending on their academic background and professional experience. We also envision the following main future work: replicate this study with other software engineering techniques, investigate how agile methodologies have been applied in practice, and detail the reasons why software metrics have been timidly applied in the industry.

\bibliographystyle{ACM-Reference-Format}
\bibliography{main}


\begin{thebibliography}{15}


\ifx \showCODEN    \undefined \def \showCODEN     #1{\unskip}     \fi
\ifx \showDOI      \undefined \def \showDOI       #1{#1}\fi
\ifx \showISBNx    \undefined \def \showISBNx     #1{\unskip}     \fi
\ifx \showISBNxiii \undefined \def \showISBNxiii  #1{\unskip}     \fi
\ifx \showISSN     \undefined \def \showISSN      #1{\unskip}     \fi
\ifx \showLCCN     \undefined \def \showLCCN      #1{\unskip}     \fi
\ifx \shownote     \undefined \def \shownote      #1{#1}          \fi
\ifx \showarticletitle \undefined \def \showarticletitle #1{#1}   \fi
\ifx \showURL      \undefined \def \showURL       {\relax}        \fi
\providecommand\bibfield[2]{#2}
\providecommand\bibinfo[2]{#2}
\providecommand\natexlab[1]{#1}
\providecommand\showeprint[2][]{arXiv:#2}

\bibitem[\protect\citeauthoryear{Bern}{Bern}{2018}]%
        {Bern2018}
\bibfield{author}{\bibinfo{person}{Baldvin~Gislason Bern}.}
  \bibinfo{year}{2018}\natexlab{}.
\newblock \showarticletitle{From Theory to Practice: Experiences of
  Industry-Academia Collaboration from a Practitioner}.
  \bibinfo{publisher}{Association for Computing Machinery},
  \bibinfo{address}{New York, NY, USA}, \bibinfo{pages}{22–23}.
\newblock
\showISBNx{9781450357449}
\urldef\tempurl%
\url{https://doi.org/10.1145/3195546.3195552}
\showDOI{\tempurl}


\bibitem[\protect\citeauthoryear{Bordin and Benitti}{Bordin and
  Benitti}{2018}]%
        {Bordin:2018}
\bibfield{author}{\bibinfo{person}{Andr{\'e}a~Sabedra Bordin} {and}
  \bibinfo{person}{Fabiane Barreto~Vavassori Benitti}.}
  \bibinfo{year}{2018}\natexlab{}.
\newblock \showarticletitle{Software Maintenance: What Do We Teach and What
  Does the Industry Practice?}. In \bibinfo{booktitle}{\emph{32th Brazilian
  Symposium on Software Engineering (SBES)}}. \bibinfo{pages}{270--279}.
\newblock


\bibitem[\protect\citeauthoryear{Canada}{Canada}{2010}]%
        {StaCan}
\bibfield{author}{\bibinfo{person}{Statistics Canada}.}
  \bibinfo{year}{2010}\natexlab{}.
\newblock \bibinfo{booktitle}{\emph{Survey Methods and Practices}}.
  Vol.~\bibinfo{volume}{Catalogue no. 12-587-X}.
\newblock \bibinfo{publisher}{Statistics Canada}.
\newblock


\bibitem[\protect\citeauthoryear{Carver, Dieste, Kraft, Lo, and
  Zimmermann}{Carver et~al\mbox{.}}{2016}]%
        {Carver:2016}
\bibfield{author}{\bibinfo{person}{Jeffrey~C. Carver}, \bibinfo{person}{Oscar
  Dieste}, \bibinfo{person}{Nicholas~A. Kraft}, \bibinfo{person}{David Lo},
  {and} \bibinfo{person}{Thomas Zimmermann}.} \bibinfo{year}{2016}\natexlab{}.
\newblock \showarticletitle{How Practitioners Perceive the Relevance of ESEM
  Research}. In \bibinfo{booktitle}{\emph{10th ACM/IEEE International Symposium
  on Empirical Software Engineering and Measurement (ESEM)}}.
  \bibinfo{pages}{56:1--56:10}.
\newblock


\bibitem[\protect\citeauthoryear{{Chidamber} and {Kemerer}}{{Chidamber} and
  {Kemerer}}{1994}]%
        {Chidamber:1994}
\bibfield{author}{\bibinfo{person}{S.~R. {Chidamber}} {and}
  \bibinfo{person}{C.~F. {Kemerer}}.} \bibinfo{year}{1994}\natexlab{}.
\newblock \showarticletitle{A metrics suite for object oriented design}.
\newblock \bibinfo{journal}{\emph{IEEE Transactions on Software Engineering}}
  \bibinfo{volume}{20}, \bibinfo{number}{6} (\bibinfo{date}{June}
  \bibinfo{year}{1994}), \bibinfo{pages}{476--493}.
\newblock


\bibitem[\protect\citeauthoryear{Fowler and Beck}{Fowler and Beck}{2018}]%
        {Fowler:}
\bibfield{author}{\bibinfo{person}{Martin Fowler} {and} \bibinfo{person}{Kent
  Beck}.} \bibinfo{year}{2018}\natexlab{}.
\newblock \bibinfo{booktitle}{\emph{Refactoring: Improving the Design of
  Existing Code} (\bibinfo{edition}{second edition} ed.)}.
\newblock \bibinfo{publisher}{Addison-Wesley Longman Publishing Co., Inc.},
  \bibinfo{address}{USA}.
\newblock


\bibitem[\protect\citeauthoryear{Kitchenham and Pfleeger}{Kitchenham and
  Pfleeger}{2008}]%
        {Kitchenham:2008}
\bibfield{author}{\bibinfo{person}{Barbara~A. Kitchenham} {and}
  \bibinfo{person}{Shari~L. Pfleeger}.} \bibinfo{year}{2008}\natexlab{}.
\newblock \bibinfo{booktitle}{\emph{Personal Opinion Surveys}}.
\newblock \bibinfo{publisher}{Springer London}, \bibinfo{address}{London},
  Chapter~3, \bibinfo{pages}{63--92}.
\newblock


\bibitem[\protect\citeauthoryear{Kupiainen, Mäntylä, and Itkonen}{Kupiainen
  et~al\mbox{.}}{2015}]%
        {Kupiainen2015}
\bibfield{author}{\bibinfo{person}{Eetu Kupiainen}, \bibinfo{person}{Mika~V.
  Mäntylä}, {and} \bibinfo{person}{Juha Itkonen}.}
  \bibinfo{year}{2015}\natexlab{}.
\newblock \showarticletitle{Using metrics in Agile and Lean Software
  Development – A systematic literature review of industrial studies}.
\newblock \bibinfo{journal}{\emph{Information and Software Technology}}
  \bibinfo{volume}{62} (\bibinfo{year}{2015}), \bibinfo{pages}{143 -- 163}.
\newblock
\showISSN{0950-5849}


\bibitem[\protect\citeauthoryear{Lo, Nagappan, and Zimmermann}{Lo
  et~al\mbox{.}}{2015}]%
        {Lo:2015}
\bibfield{author}{\bibinfo{person}{David Lo}, \bibinfo{person}{Nachiappan
  Nagappan}, {and} \bibinfo{person}{Thomas Zimmermann}.}
  \bibinfo{year}{2015}\natexlab{}.
\newblock \showarticletitle{How Practitioners Perceive the Relevance of
  Software Engineering Research}. In \bibinfo{booktitle}{\emph{10th Joint
  Meeting on Foundations of Software Engineering (FSE)}}.
  \bibinfo{pages}{415--425}.
\newblock


\bibitem[\protect\citeauthoryear{Murphy-Hill, Parnin, and Black}{Murphy-Hill
  et~al\mbox{.}}{2012}]%
        {MurphyHill2012}
\bibfield{author}{\bibinfo{person}{Emerson Murphy-Hill}, \bibinfo{person}{Chris
  Parnin}, {and} \bibinfo{person}{Andrew~P. Black}.}
  \bibinfo{year}{2012}\natexlab{}.
\newblock \showarticletitle{How We Refactor, and How We Know It}.
\newblock \bibinfo{journal}{\emph{IEEE Trans. Softw. Eng.}}
  \bibinfo{volume}{38}, \bibinfo{number}{1} (\bibinfo{date}{Jan.}
  \bibinfo{year}{2012}), \bibinfo{pages}{5–18}.
\newblock
\showISSN{0098-5589}
\urldef\tempurl%
\url{https://doi.org/10.1109/TSE.2011.41}
\showDOI{\tempurl}


\bibitem[\protect\citeauthoryear{{Palomba}, {Bavota}, {Di Penta}, {Oliveto},
  and {De Lucia}}{{Palomba} et~al\mbox{.}}{2014}]%
        {Palomba2014}
\bibfield{author}{\bibinfo{person}{F. {Palomba}}, \bibinfo{person}{G.
  {Bavota}}, \bibinfo{person}{M. {Di Penta}}, \bibinfo{person}{R. {Oliveto}},
  {and} \bibinfo{person}{A. {De Lucia}}.} \bibinfo{year}{2014}\natexlab{}.
\newblock \showarticletitle{Do They Really Smell Bad? A Study on Developers'
  Perception of Bad Code Smells}. In \bibinfo{booktitle}{\emph{2014 IEEE
  International Conference on Software Maintenance and Evolution}}.
  \bibinfo{pages}{101--110}.
\newblock


\bibitem[\protect\citeauthoryear{{Parnas}}{{Parnas}}{2011}]%
        {Parnas:2011}
\bibfield{author}{\bibinfo{person}{D. {Parnas}}.}
  \bibinfo{year}{2011}\natexlab{}.
\newblock \showarticletitle{Software Engineering - Missing in Action: A
  Personal Perspective}.
\newblock \bibinfo{journal}{\emph{Computer}} \bibinfo{volume}{44},
  \bibinfo{number}{10} (\bibinfo{date}{Oct} \bibinfo{year}{2011}),
  \bibinfo{pages}{54--58}.
\newblock


\bibitem[\protect\citeauthoryear{Runeson}{Runeson}{2012}]%
        {Runeson2012}
\bibfield{author}{\bibinfo{person}{Per Runeson}.}
  \bibinfo{year}{2012}\natexlab{}.
\newblock \showarticletitle{It Takes Two to Tango -- An Experience Report on
  Industry -- Academia Collaboration}. In \bibinfo{booktitle}{\emph{Proceedings
  of the 2012 IEEE Fifth International Conference on Software Testing,
  Verification and Validation}}. \bibinfo{pages}{872–877}.
\newblock
\urldef\tempurl%
\url{https://doi.org/10.1109/ICST.2012.190}
\showDOI{\tempurl}


\bibitem[\protect\citeauthoryear{Society, Bourque, and Fairley}{Society
  et~al\mbox{.}}{2004}]%
        {ieee2004_guide}
\bibfield{author}{\bibinfo{person}{IEEE~Computer Society},
  \bibinfo{person}{Pierre Bourque}, {and} \bibinfo{person}{Richard~E.
  Fairley}.} \bibinfo{year}{2004}\natexlab{}.
\newblock \bibinfo{booktitle}{\emph{Guide to the Software Engineering Body of
  Knowledge (SWEBOK)}}.
\newblock \bibinfo{publisher}{IEEE Computer Society Press}.
\newblock


\bibitem[\protect\citeauthoryear{Tahir, Yamashita, Licorish, Dietrich, and
  Counsell}{Tahir et~al\mbox{.}}{2018}]%
        {Amjed2018}
\bibfield{author}{\bibinfo{person}{Amjed Tahir}, \bibinfo{person}{Aiko
  Yamashita}, \bibinfo{person}{Sherlock Licorish}, \bibinfo{person}{Jens
  Dietrich}, {and} \bibinfo{person}{Steve Counsell}.}
  \bibinfo{year}{2018}\natexlab{}.
\newblock \showarticletitle{Can You Tell Me If It Smells? A Study on How
  Developers Discuss Code Smells and Anti-Patterns in Stack Overflow}. In
  \bibinfo{booktitle}{\emph{Proceedings of the 22nd International Conference on
  Evaluation and Assessment in Software Engineering 2018}}.
  \bibinfo{pages}{68–78}.
\newblock
\urldef\tempurl%
\url{https://doi.org/10.1145/3210459.3210466}
\showDOI{\tempurl}


\end{thebibliography}


\end{document}